# Homodyne QPSK Detection for Quantum Key Distribution

**M. B. Costa e Silva, Q. Xu, S. Agnolini, and P. Gallion**
*Ecole Nationale Supérieure des Télécommunications (GET/Télécom Paris and CNRS),*
*46 rue Barrault, 75013 Paris, France.*
Email: mcosta@enst.fr

**F. J. Mendieta**
*CICESE, km.107 Carr. Tijuana, Ensenada, Baja California 22800, México.*

**Abstract:** We present a QKD system with fainted pulses using self-homodyne coherent detection in optical fibers at 1543nm. BB84 protocol key is encoded in the optical phase using a two-electrode Mach-Zehnder modulator, producing a QPSK modulation.
@2006 Optical Society of America
**OCIS codes:** (060.2920) Homodyning; (060.5060) Phase Modulation; (999.9999) Quantum-Key-Distribution

## 1. Introduction.

In fiber-optic quantum key distribution (QKD) at telecommunications wavelengths, coherent homodyne detection of phase encoded keys constitutes an attractive alternative to APD photon counters, which suffer from low quantum efficiency and long response time due to the quenching process [1].

In the search for higher key rates, coherent homodyne detection of these formats provides additional advantages such as the operation near the quantum limit, with the mixing gain to combat the thermal noise, obtained with standard p.i.n. detectors, which exhibit in general less noise sensitivity, higher quantum efficiency, faster response, as well as lower cost and power requirements than quenched avalanche photodiodes [2].

When a strong, phase and polarization synchronized reference is regenerated, a variety of key encoding formats is detectable; in this work we present two configurations: a self-homodyne, and a delayed homodyne scheme that uses only one fiber where a reference is time-multiplexed with the encoded key.

## 2. Experimental Setup

In our BB84-QKD phase modulated system, Alice encodes its symbols as antipodal phase states in two conjugated bases, resulting in a QPSK format. We implement this modulation by using a two-electrode Mach-Zehnder electro-optical modulator (EOM-A), with the pre-modulation arrangement as shown in Fig. 1 [3], allowing the independent choice of base and symbol as required by BB84, producing the following Alice's QPSK transmitted field:

$$E_A(t) = E_{A_0}(t) \cdot \exp\left(j\left[(\Phi_1 + \Phi_2)/2\right]\right) \qquad (1)$$

In order to maintain the constant envelope condition, the intensity modulation term must be kept unchanged, so we construct the Alice's encoding table as shown also on Fig. 1, where we have indicated the Bob's choice of base, producing the following optical field resulting from Bob's measurement by applying its phase modulation with a similar modulator EOM-B, using one electrode only:

$$E_B(t) = E_{B_0}(\Phi_B) \cdot \exp\left(j(\Phi_A - \Phi_B)\right) \qquad (2)$$

When Alice's and Bob's bases coincide ($\Phi_A - \Phi_B = 0$ or $\pi$), the phase takes one of two antipodal values that can be measured, but in the case of anti-coincidence ($\Phi_A - \Phi_B = \pi/2$ or $-\pi/2$), the measures are random and non informative.

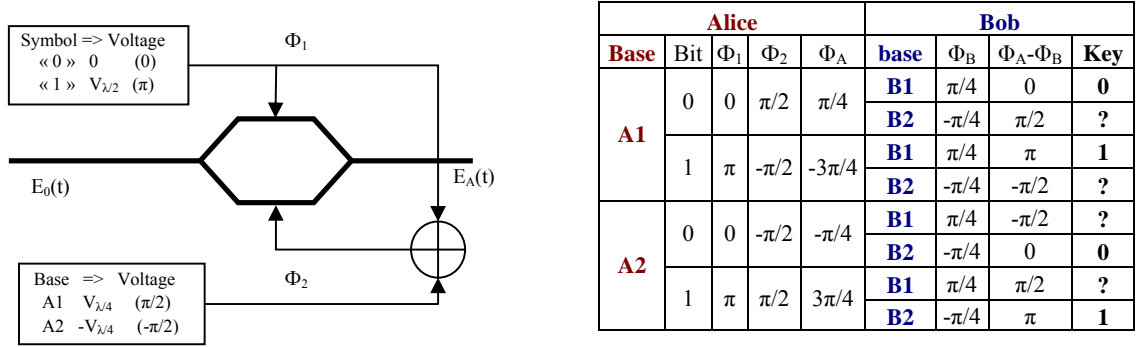

Fig. 1 Alice's encoding scheme and table for QPSK protocol BB84

*2.1 Self-Homodyne configuration*

Our first setup consists of a standard telecommunications fiber-optic self-homodyne with the strong carrier "reference" transmitted in a separate line. In the "signal" line Alice introduces her base and symbol choices by phase modulating the EOM-A, and strongly attenuates the signal before sending it to Bob, who also applies his 2-state phase modulation as base's choice. Bob performs then homodyne photodetection with 2 p.i.n. photodiodes in a balanced configuration. Fig. 2(a) shows a typical detected output when a pulse modulated laser is sent to the system, for a received power of -47dBm. The coincidence of bases between Alice and Bob are shown in the waveform as positive and negative pulses, while the pulses corresponding to anti-coincidence are discarded. Fig. 2(b) shows a histogram of the detected output (as shown in Fig. 2(a)), where we can observe the manifestation of the base coincidence (outer peaks) and anti-coincidence (inner peak).

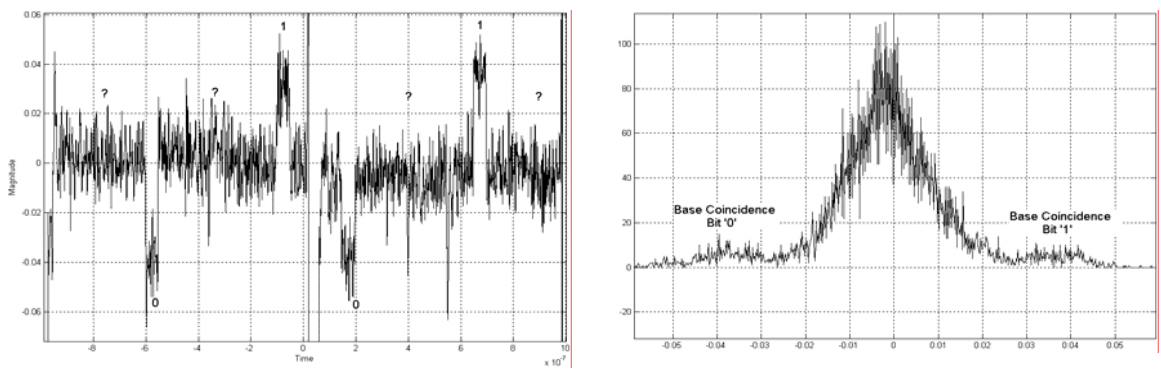

Fig. 2 (a) Detected Output and (b) Histogram for Pr = -47dBm

*2.2 Delayed homodyne configuration*

The first self-homodyne configuration requires the transmission in two fibers, which suffer from unequal phase and polarization shifts and stabilities as the propagation distance increases. In addition, the "signal" and "reference" line must be in absolute phase-alignment in order to reach a suitable self-homodyne configuration, which is very difficult to implement in a practical application.

Our second homodyne configuration consists in sending the weak QPSK-modulated pulses time-multiplexed with unmodulated strong pulses that constitute a carrier phase reference in a same fiber. Fig. 3 is a diagram of our experimental setup: optical pulses are fed into Alice's unbalanced interferometer: Alice's fainted pulse QPSK signal is produced in its longer arm with an EOM-A as mentioned above, while strong unmodulated pulses pass through the shorter arm, with accurate polarization control.

At the receiver, Bob's measurements are performed by applying his 2-state phase modulation to the strong pulses in a similar delayed interferometric configuration so that the weak key pulses beat with the high power reference pulses, in order to achieve an acceptable mixing gain; then a balanced photodetection is performed. Fig. 4(a) shows the detected output after 11 km fiber propagation and 4(b) the corresponding histogram.

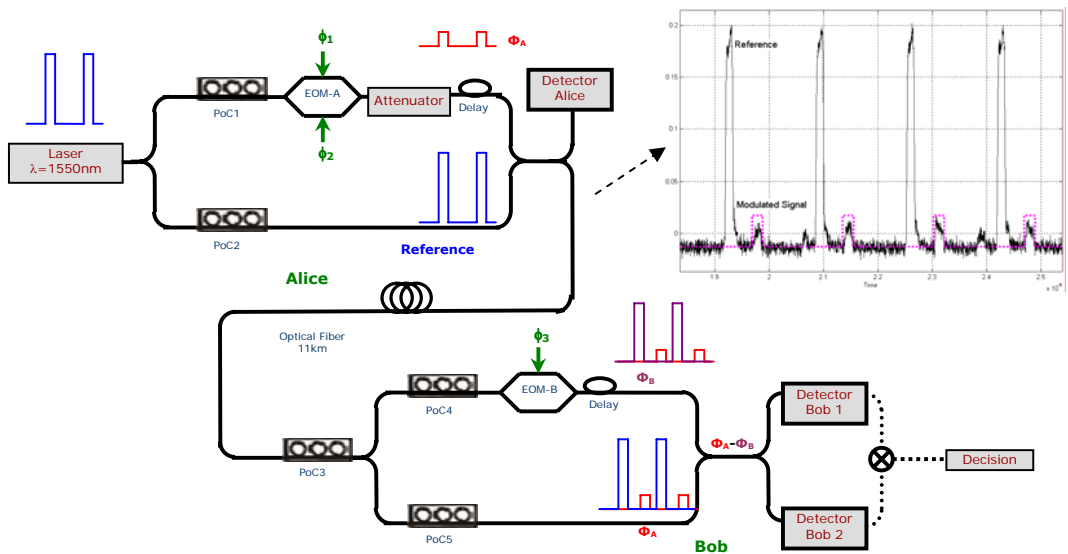

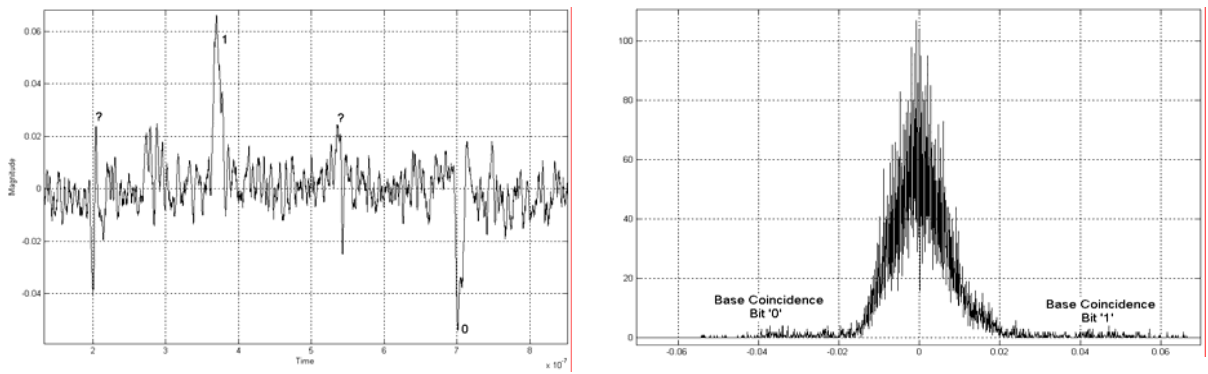

Fig.3 Delayed homodyne QKD Setup

Fig. 4 (a) Detected Output and (b) Histogram after 11 km fiber propagation

### 3. Conclusion

We have implemented a fiber-optic coherent homodyne detection system for QPSK modulation scheme in QKD applications using standard telecommunications fibers, which employs p.i.n. photodetectors as an alternative to APD photon counters. We introduced a technique for independent choice of Alice's base and symbol. We present results on detected outputs and statistics on base coincidence and anticoincidence for two configurations: a self-homodyne and a delayed homodyne scheme. The second scheme reduces impressively the phase stability requirements due to the transmission in the single fiber and provides a much easier and stabler long distance QKD system.